# Void superlattice formation as self-organization phenomemon. 1. Scaling estimates

V.N. Kuzovkov[1], E.A. Kotomin[1*], G.Zvejnieks[1], K.D. Li[2], L.M. Wang[2]

[1]Institute for Solid State Physics, University of Latvia, 8 Kengaraga, LV - 1063 Riga, Latvia

[2]Dept. Materials Science & Engineering, and Department of Nuclear Engineering & Radiological Sciences, University of Michigan, Ann Arbor, Michigan 48109, USA

**Abstract**

The literature on void superlattice formation observed under irradation of metals and insulators is analyzed with a special attention to the self-organization; a general scenario of this process is discussed. A simple relation for the superlattice parameter as a function of the dose rate and temperature is suggested, in a good agreement with existing experimental data. Special attention is paid to analysis of the halogen gas void formation in an electron-irradiated $CaF_2$.

(version January 22, 2009)

## I. INTRODUCTION

As is well known, irradiation of many metallic and insulating solids with energetic particles, such as heavy ions, neutrons, electrons, can result in a formation of *ordered structures* including periodic defect walls, bubble lattices, void lattices and periodic compositions in alloys [1-3]. The particular ordered structures arising in such *open dissipative systems far from equilibrium* depend on a type, energy, and flux of the energetic particles as well on the temperature. It was noticed [3] that despite the difference in the appearance, a similar underlying mechanism may be invoked to explain the *self-organization behavior* of these structures. In this paper, we focus on the *void superlattice formation* in which a long-range ordered superlattice is created with the lattice parameter of the order of 50 nm. Such superlattices are relatively well studied experimentally in both metals and insulators [1-3].

The very fact that this is a self-organization process poses considerable limitations on the theoretical methods which could be used for its treatment and calls in question many void formation scenarios suggested so far in the literature (see e.g. review article [2]). The reference [3] could be mentioned as an illustration of this point, where the mechanism of a spinodal decomposition in a radiation-induced pattern formation was discussed. However, this approach implies the temporal and spatial evolution of the system towards thermodynamic equilibrium. The more so, the interstitial atoms were neglected and thus the void concentration remains constant (no vacancy recombination with interstitials). However, this has nothing to do with open dissipative systems far from equilibrium under study. It contradicts also to the experimental fact that the voids grow and start ordering under continuous irradiation [3].

A similar criticism could be applied to the attempts to determine the distinctive lattice parameter of the superlattice assuming that this is a space scale characterizing the local minimum in a potential energy curve (quasi-elastic analysis) [4-6]. As noticed in Ref. [7], "while void-void elastic interaction is strong when voids are closely separated, the short-range of the elastic interaction forces does not explain how voids and bubbles organize themselves over relatively long distances, especially during the early stages of irradiation". Thus, no surprise that in these theories, however, the energy minima rarely correspond to the observed void lattice parameters or symmetries. The key point here is that the superlattice parameter characterizes the non-equilibrium process. This has also been shown by a recent model, which reveals the dynamic void formation process under the strain field induced by the surface stress at the void/solid interface [3]. A good correlation was established between the nonequilibrium superlattice structures and experimental observations. It was also shown that elastic anisotropy can significantly influence the symmetry of a void superlattice, causing it to replicate that of the host crystal.

The self-organization process leads to the two observations which at the first glance look contradicting. On the one hand, when void lattice is observed, its properties are surprisingly insensitive to the external conditions, such as material, temperature, irradiation type. In particular, it was observed in metals that the lattice parameter of the void lattice is largely insensitive to defect production rates, dose (for high doses) and details of the particle recoil spectrum [1,7]. The void lattice parameter decreases slightly with increasing damage rate but increases slightly with increasing irradiation temperature over a large range [1,7]. On the other hand, this is very unstable effect: "The phenomenon was only observed in a very small number of samples within a narrow condition window. Sometimes the results cannot be repeated presumably due to small variations of the experimental conditions that were thought unimportant before" [8]. This can lead to the hasty conclusion that it is indeed very true that the irradiation conditions seem to be in a very narrow window [8]. This contradiction is resolved if we take into account that that self-organization phenomena are not deterministic, the *reactant density fluctuations* play a great role here [9].

This could be well illustrated by recent studies for other self-organized systems, e.g. the spatio-temporal oscillations in catalytic CO oxidation on Pt surface [10]. The nonlinear kinetics of surface reactions therein shows a variety of phenomena such as many kinds of pattern formation, global oscillations, and even chaotic behavior. The ordered structures (reactant patterns) were observed [10] along with the labyrinth-like structures, which looks similar to radiation-induced well aligned labyrinth-like defect walls [3]. The different-type spatio-temporal structures compete each other. In particular, it was observed in the mentioned CO oxidation study by means of the Monte Carlo modeling that the long-time asymptotics (the structure type) was determined by a *random parameter b* —the difference in a number of spirals with opposite rotation directions which arise in the beginning of a non-linear process. For $b = 0$ the opposite-type spirals annihilate in turn each other, due to which other structures can arise, e.g. well-ordered global synchronization of reactant densities [10]. However, even in this case there is no distinctive time of the structure formation which, in fact, is a random parameter. These observations are relevant also for self-organization under irradiation. Along with the void superlattice, other above-mentioned ordered structures can arise and compete each other, and the choice between these is a random process. This is why any attempts to detect the *parameter window* for the void lattice observation hardly could be successful.

In Section 2 we discuss in a more detail several possible stages of a self-organized process of void lattice formation based on the analysis of existing experimental literature. In Section 3 general scaling estimates for void growth are performed. These estimates are applied for metals and insulators under electronic irradiation ($CaF_2$) in Section 4. The conclusions are summarized in Section 5.

## 2. Three stages of void lattice formation
### 2.1. General consideration

The void superlattice formation could be divided into three stages. At the first stage, voids initially are formed randomly in irradiaded materials [7]. This is true also for other patterns, e.g. periodic defect cluster walls [1] where the ordered cluster arrangements are formed from an initially random cluster distribution at low doses; as well as gas bubble lattices. In other words, at low doses first of all, disordered structure arises which elements (voids) are large and stable enough to survive coalescence or annihilation with interstitials. The voids are similar to atoms in disordered system such as glass or liquid. The traditional general statement that voids initially are formed randomly in irradiaded materials should be completed with several fundamental void lattice properties [7]:

(i) The random void structure parameter is typically about two orders of magnitude larger than the atomic lattice parameter. (ii) The random void structure parameter decreases slightly with increasing damage rate. (iii) The random void structure parameter increases with increasing the irradiation temperature. Peculiarities of the disordered void structure determine the further stages of the process.

At the second stage, this structure has a trend to "crystallization" through formation, first a short-order and then a long- (global-) order. According to ref. [1,8], the degree of ordering improves with increasing the irradiation dose. Migration and preferential growth were dominant at the final stages of the superlattice formation [7].

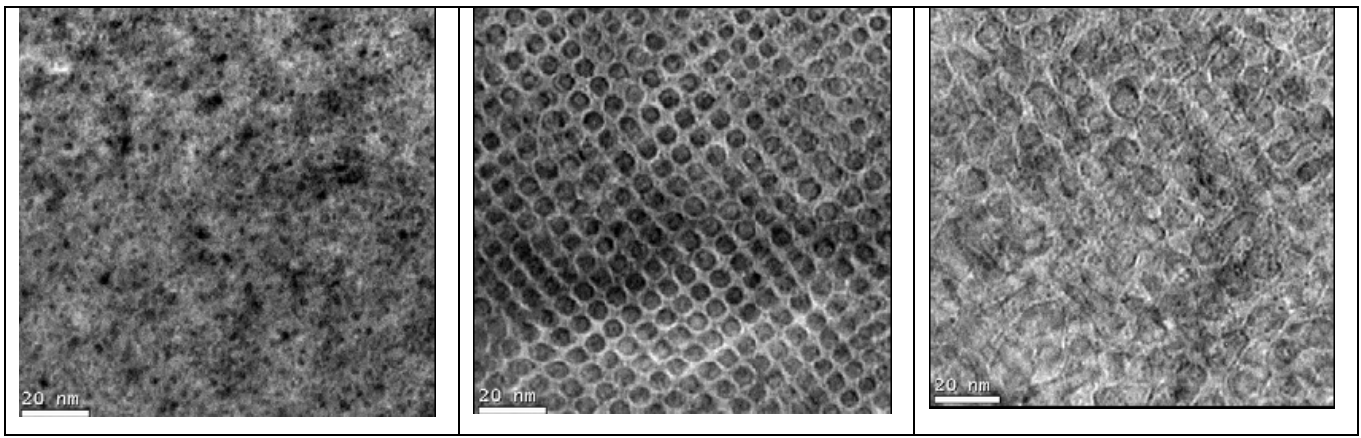

Fig. 1. Three stages of void lattice formation in $CaF_2$ [8].

Two statements could be found in the experimental literature: (a) The void lattice parameter decreases slightly with increasing damage rate [7], and (b) the void ordering is insensitive to the dose rate [3]. The first statement corresponds to the first stage, whereas the second statement could be associated with the second stage of the system ordering because the mean distances between voids in disordered and ordered structures are close. Independence on the dose rate means that a permanent irradiation is the necessary condition for existence of the glass structure but irradiation itself does not order the void structure. The individual voids reveal a continuous growth without coalescence but the superstructure lattice parameter remains constant.

A variation of the void radius leads to the final, third stage of a self-organization process: at a critical dose the superlattice structure is destroyed. These critical doses of void superlattice formation and deformation seem to be independent of dose rate [8]. Similar observation was made for the gas bubble lattices [1]: "At extremely high doses ordering disappears and, instead, a network of bubbles and channels without ordering develops". Again, the process here, similarly to the stage 2, is not sensitive to the dose rate.

The processes of ordering at the second stage and disordering at the third stage possess different rates. These are diffusion-controlled processes involving migration of individual vacancies and interstitials at the first stage and voids at the second stage. Existence of three stages of self-organization indicates that the void superlattice formation is not a steady-state but an *intermediate* asymptotics [9] observed in a limited range of the time and the radiation doses. As we show below, the superlattice parameters could be estimated already at the first stage of the process.

### 2.2. Previous theoretical studies

Several kinetic studies were undertaken to simulate theoretically the void formation. Fist of all, these are Monte Carlo studies [11,12]. Instead of modelling *the kinetics* of void formation and growth, authors [11] started from a random array of small voids and introduced randomly positioned 1d crowdions and vacancy clusters that interact with the voids. The more so, such an important radiation effect as void nucleation has been omitted. As a result, the radiation-damage problem is transformed into relaxation kinetics of SIA-clusters (self-interstitial atoms) and vacancy recombination. Since the void concentration decays due to irreversible reaction, the scenario used was to start with a high enough concentration of small voids, very near every position that a lattice void will occupy [11]. The reasonable question was raised in Ref.[12], whether it is reasonable in the face of strong coalescence loss of voids to ask what the effect of continued void nucleation might be. Evans [12] tried to overcome limitations of the modeling [11] through the effects of renucleation and the influence of vacancies. However, this is done in a way very far from real process occurring under irradiation. In fact, in his simulations any void lost due to shrinkage or coalescence was replaced by a new void having the original starting radius. The new void was given random coordinates. As a result, the formation of a perfect superlattice seemed to be elusive. As the author concluded, "there is no indication in the present work that the almost perfect void lattices or bubble lattices that have been produced experimentally could be a result of 1d SIA transport" [12]. The more so, 1d SIA motion is not the case for insulators where void lattice was also observed.

Another series of kinetic papers [3,7,9,13-15] was based on the standard mesoscopic self-organization approach. This assumes some intermediate steady-state with *homogeneously* distributed reactants (e.g. considerable steady-state concentration of vacancies and interstitials). In other words, it is assumed that under continues irradiation high concentration of *single* defects is created but no voids or SIA clusters. Then stability of this state is considered with respect to a small perturbation characterized by the wavenumber $k$. Mathematically this means a *bifurcation analysis* of the non-linear differential equations. As a result, spatially-inhomogeneous periodic solution could be obtained with the superstructure $L = 2\pi/k_0$. In fact, this contradicts our kinetic Monte Carlo modelling [17] indicating at similar defect aggregation (void formation) from the very beginning of the irradiation process. Thus, this mesoscopic approach is unable to predict the kinetics of the radiation damage accumulation and time-development of the void system evolution. The more so, this is a mean-field theory where reactant density fluctuations are, in fact, neglected. We show in Section 3 that the first stage of the void formation cannot be treated in terms of the mean-field theory, in particular, because it neglects formation of defect clustering at early stages of radiation damage and studies unstabilities of a homogeneous defect distribution at quite high doses.

Lastly, Fokker-Planck-type kinetic equations were applied [16] in order to study effects of diffusion anisotropy and one-dimensional motion of crowdions in metals. This approach also does not take into account strongly non-equilibrium nature of the system and focused on the specific situation in metals.

## 3. Scaling theory of void structure
### 3.1. Preliminary estimates

Let us make simple estimates of vacancy (v) and interstial (i) aggregation driven by their diffusion with the coefficients $D_v$ and $D_i$. The standard diffusion coefficient $D = D_0 \exp(-E_a/k_B T)$, where $E_a$ is the migration energy. Typically $D_0 \sim 10^{-3}$ cm$^2$/s and the migration energy for vacancies $E_a(v)$ in both metals and insulators considerably exceeds that for interstitials, $E_a(i)$ (e.g. 1.3 eV and 0.3 eV for Ni [7] or 1.0 eV and 0.1 eV for NaCl ([15], respectively and references therein). The diffusion coefficient could be written also as $D = \nu a_0^2$ where $\nu$ is the jump frequency, $a_0$ the lattice constant (typically ~ 4 Å). The temperature range of interest $T$ = 300-1000 K.

Simple qualitative estimate of jump frequencies for single mobile vacancies and interstitials shown in Fig.2 demonstrates the orders of magnitude difference in defect mobilities which makes direct 3d Monte Carlo modelling unrealistic: since the time step is defined by very mobile interstitials, it is hardly possible to study vacancy system evolution over reasonable period of time. In such systems the scaling estimates could be the first step.

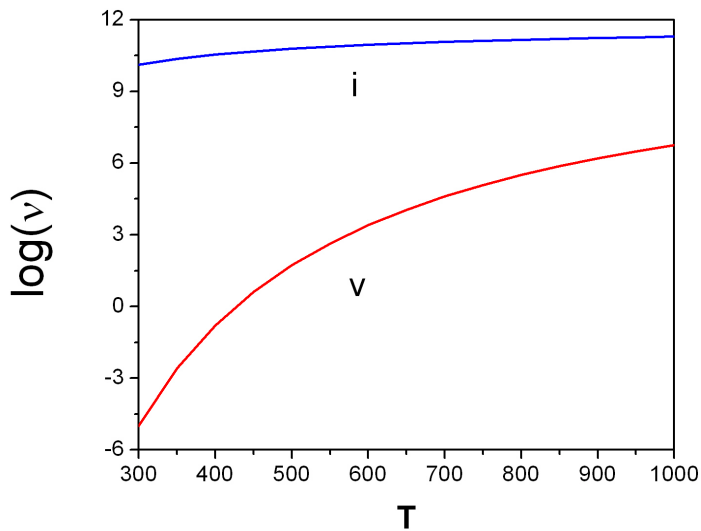

Fig.2. The jump frequency of vacancies and interstitials (in logarithmic scale) as a function of temperature for $E_a(i) = 0.1$ eV and $E_a(v) = 1$ eV.

Typical experimental dose rates $p$ vary (dependent on irradiation type and material) in the range of $p = 10^{-6} - 10^{-3}$ dpa/s. The threshold dose $G$ for the void lattice formation in b.c.c. metals is a few dpa (e.g. [7], in f.c.c. metals it is larger by an order of magnitude. Thus, for estimates we use the dose $G = pt = 10$ dpa. The superlattice parameter ranges $L = 200 - 1500$ Å, whereas the void diameter $L_0$ is comparable with the $L$ [1,2,7]. There is a clear correlation between $L$ and $L_0$ (Fig.3 in Ref. 2 and Fig.3 in this paper): typically $L_0 / L \sim 0.25$. It is convenient to use hereafter dimensionless lattice parameter $\lambda$ and void diameter $\lambda_0$ defined as $L = a_0 \lambda$, and $L = a_0 \lambda_0$.

We assume that voids are dense agglomerates of vacancies. The dimesionless void concentration could be estimated as $C_v = (\lambda_0 / \lambda)^3 = 10^{-2}$, which is comparable with an estimate of a maximum possible concentration of accumulated *immobile* defects $C_v \sim 0.1$ [9]. For a typical $G = 10$ dpa, 10 defects are created in each unit cell and only 0.1 % of the totally produced defects survive [1] due to a vacancy-interstitial annihilation. If free interstitial atoms were presented in the same concentration, they would definitely destroy a void structure. Indeed, in random walks with diffusion coefficient $D$ a particle during time $t_R$ coveres the distance $R$ defined as $t_R = R^2 / D$. For the typical parameters cited above the interstitials would collide with voids every $10^{-6}$s. This time could be compared with the irradiation time of the order of $10^4$-$10^7$s, their ratio is astronomically large: $10^{10}$-$10^{13}$! That is, obvious conclusion could be drawn that free and highly mobile interstitials should be bound or trapped somewhere.

It is also hardly possible that interstitials segregate to the surface since similar estimate shows that while segregationg to a 1 cm thick sample surface interstitials have to collide $10^{10}$ times with voids and would definitely annihilate. It is commonly believed (e.g. [7]) that interstitials disappear to the immobile dislocation loops. However, there is no clear quantitative information about density of interstitials in dislocations and even more important - how aperiodic dislocation distribution affects the periodic void superlattice. Another above-mentioned option is that interstitials are also bound into immobile aggregates similar to the voids. Let us make estimates for such process.

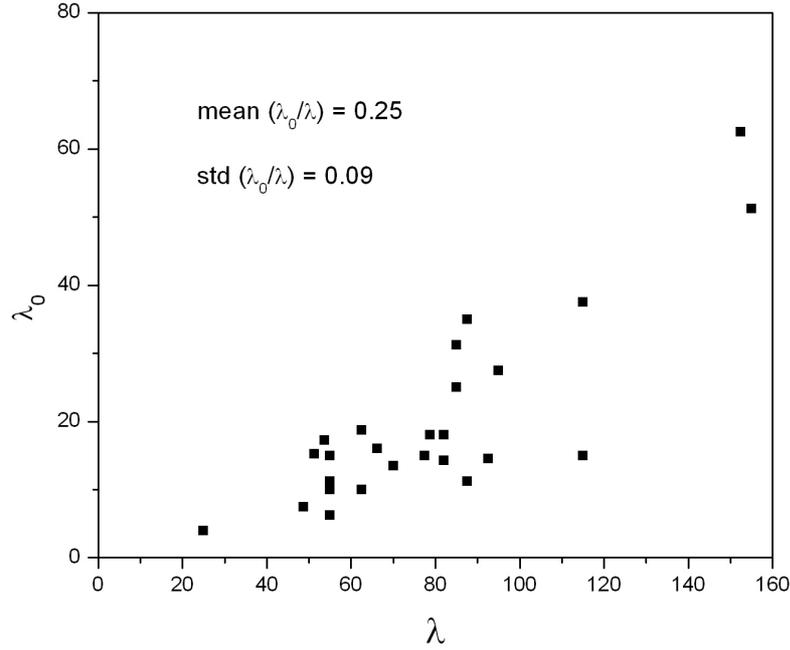

Fig. 3. Dependence of the observed dimensionless void diameter $\lambda_0$ as a function of the dimensionless void superlattice constant $\lambda$ for a number of metals (based on data from Table 2 [7]). The calculated mean value of the ratio $\lambda_0 / \lambda$ is 0.25, std means the standard deviation.

### 3.2. Scaling estimates

Let us start with a simple model: single-type particles are created with the dose rate $p$, perform random walks with the diffusion coefficient $D$ (a jump frequency $\nu = D a_0^{-2}$) and form immobile aggregates when encounter each other. At low doses, $G = p t_{max} \ll 1$, the aggregate overlap could be neglected. We expect qualitatively that a disordered cluster system with a distinctive spatial parameter $\xi$ is developed. (This is supported by our Monte Carlo modeling for a similar surface problem [17] and more refined 2d MC modeling with defect annihilation [18].) Formation of such a system is a random process, the primary dimer germs are created at arbitrary coordinates where two similar particles meet and become immobile.

In the limiting case of strongly bound aggregates any just created mobile particle has a short lifetime before it finds another particle or aggregate and becomes immobile. This lifetimes is (by an order of magnitude) $t_0 = \xi^2 / \nu$ where $\xi$ is an average dimensionless distance between immobile aggregates. At low dose rates concentrations of free particles is low and growth of existing aggregates dominate over formation of new small aggregates. This is true if in the volume $V = \xi^d$ covered by a newly created particle during time $t_0$ only one particle is created, $pVt_0 = 1$ ($d$ is a space dimension). From these two relations the characteristic distance between voids- the *diffusion length* $\xi$ – could be easily obtained (cf [19])

$$\xi = (\nu/p)^{1/(d+2)} \tag{1}$$

Detailed analysis of Eq.(1) is discussed below. The aggregate diameter $\xi_0$ at arbitrary time $t$ could be estimated as a fraction of defect-occupied volume which approximately equals to the dimensionless dose, $(\xi_0/\xi)^d = pt = G$. The critical dose at which the self-supported system start to disappear due to aggregate overlap is $G_c \sim 0.1$.

Let us consider now the case of the *two* types of particles – vacancies and interstitials – which can annihilate with each other or create the aggregates of dissimilar particles. (This is the case of the electron irradiation of insulators [8]). It is shown in the kinetic MC modeling [17] that the two subsystems with two relevant spatial parameters for voids ($v$) and interstitial ($i$) aggregates are formed :

$$\xi_{v,i} = (\nu_{v,i}/p)^{1/5} \tag{2}$$

A small power factor 1/5 arises here for a real case of d=3.
It should be noted that similarly to the one-component system, the preferential growth of both voids and interstitial clusters remains but their annihilation reduces the aggregate growth

$$(\xi_{0v}/\xi_v)^3 = (\xi_{0i}/\xi_i)^3 = f(pt), \tag{3}$$

Where $f(pt)$ is a slowly increasing function of time, $\xi_{0v,i}$ are diameters of the corresponding i- and v- aggregates.
An important conclusion arising from Eq. (3) is that

$$(\xi_{0v}/\xi_{0i}) = (\xi_v/\xi_i), \tag{4}$$

i.e. the ratio of the i- and v-aggregate radii equals to that of the relevant superstructure parameters. Summing up, the dependence of the diffusion length $\xi$ for interstitials and vacancies as a function of the dose rate and temperature is predicted by Eq. (2) and illustrated in Fig. 4 for the typical metal parameters.

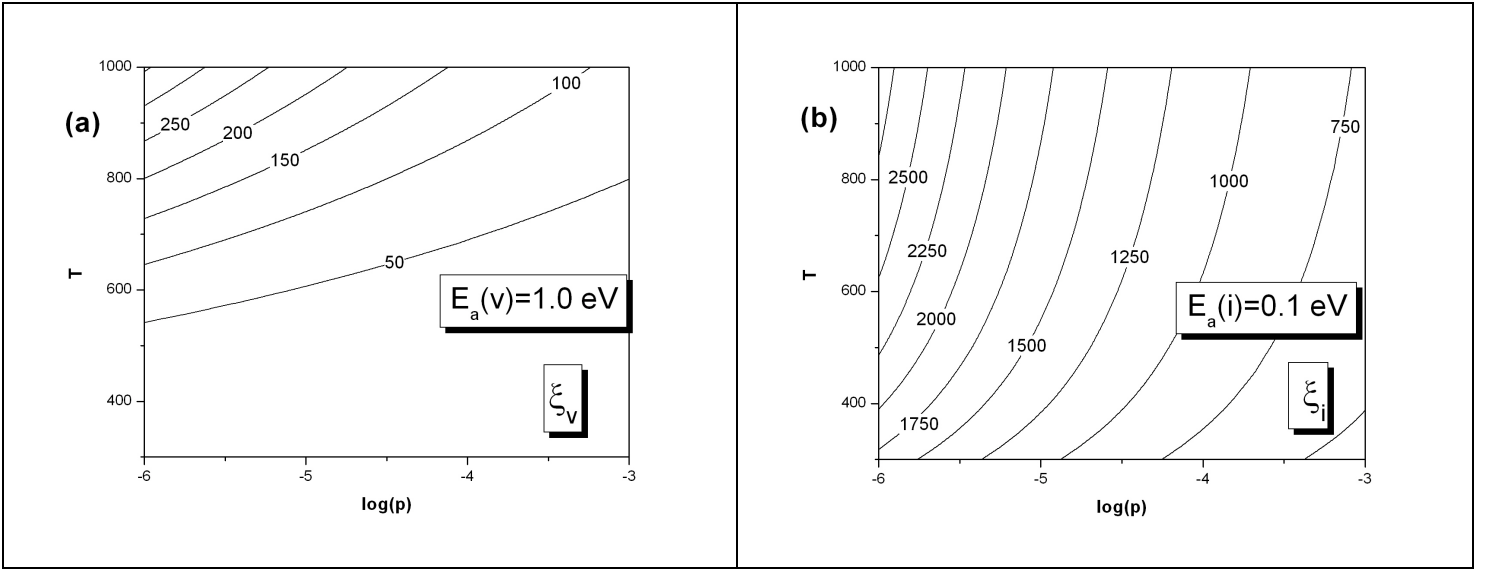

Figure 4: The predicted distinctive diffusion lengths $\xi$ (in units of $a_0$) for a self-organization of vacancy clusters (a) and interstitial aggregates (b) as a function of the dose rate $p$ and the temperature $T$ for the typical migration energies: $E_a(i) = 0.1$ eV, $E_a(v) = 1$ eV. The diffusion energies typical for metals are given in a legend.

### 4. Analysis of experimental data
### 4.1. Metals

Our predictions (Eq. (2) and Fig. 4) are in a good agreement with three basic experimental observations for vacancy void lattices in metals (e.g.[2,7] and Fig. 5): (i) the diffusion length decreases with increasing the dose rate; (ii) it increases with the temperature; and (iii) the diffusion length is typically about two orders of magnitude larger than the perfect lattice parameter. We explain also a weak dependence on the dose rate $p$ and jump frequency $\nu$ by the power factor $1/(d+2)=1/5$ in Eq.2.

Such an excellent agreement of our predicted behaviour for the diffusion length (short-range parameter) $\xi_v$ and the experimental void lattice (long-range) parameter $\lambda$ (Fig. 5) permits us to suggest their identity $\xi_v \sim \lambda$. This indicates that at the first stage of above-described self-organized aggregation process voids are created in a disordered system with the distinctive mutual distance $\xi_v$. When the ordering occurs, the system "density" (void number per volume) does not change considerably and the superlattice is formed with the lattice constant $\lambda$ close to the $\xi_v$.

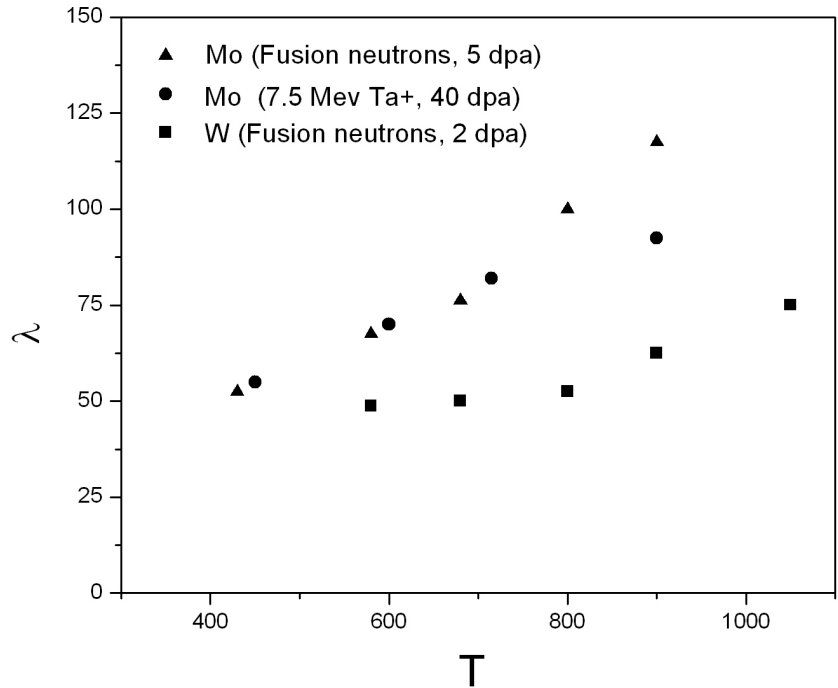

Fig.5. The temperature dependence of the void superlattice constant for metals (based on data in Table 2 [7]).

As to the interstitials, it is generally believed that in metals their preferential adsorption dy dislocations due to a stronger eleastic interaction leads to dislocation climb formation and introduces a bias in the defect fluxes to the sinks [1,20]. This is not the case, however, for insulators (see next Section 4.2).

### 4.2. The electron irradiation of $CaF_2$

It is well known [21] that the excitonic mechanism of the radiation damage of MeX insulators qualitatively differs from that in metals: isolated anion Frenkel pairs are produced instead of displacement cascades. At low doses anion atom X moves to the interstitial position thus forming the so-called *H* center and leaves a vacancy with trapped electron behind (called the *F* center). Cation sublattice remains practically undamaged. These defects are paramagnetic and well observed by means of ESR. At moderate and high teperatures these two types of defects start to migrate and aggregate. When a cluster of the *F* centers is created with Me ions inside, a system collapses into a colloid consisting of tens or hundreds of metal atoms. It was indeed well observed that under prolonged irradiation metal colloids and gas bubbles are developed (e.g. in NaCl [22]) which could annihilate in a back reaction. The void superlattice in the electron irradiated $CaF_2$ was observed recently [8,23] which differs considerably from the pattern typical for metals: the lattice parameter is very small, $\lambda \sim 50$, whereas void diameter $\lambda_0$ is very large, so that $\lambda_0 / \lambda \sim 0.5$.

For this system we can apply the model developed above in Section 3 assuming Ca ions remain immobile. There is considerably uncertainty about migration energies for vacancies and interstitials in $CaF_2$. Due to a close packing of the fluorine ions in the <100> direction, the relevant vacancy migration energy is quite low, ~0.33 eV [24,25] whereas in other directions it is much higher, > 2 eV. However, the kinetics of defect recombination, aggregation and finally void formation is controlled by 3D diffusion which difficult to estimate from such theoretical data. This is why we used here the experimental estimes based on the kinetics of *F-H* center recombination [26] $E_a(i) = 0.4$ eV, $E_a(v) = 0.7$. These parameters were used by us earlier in the successful modeling of the metal colloid formation in $CaF_2$ irradiated by low-energy electrons [27]. Typical dose rates of the electron irradiation correspond to the range of $p = 10^{-4} - 10^{-1}$ dpa/s.

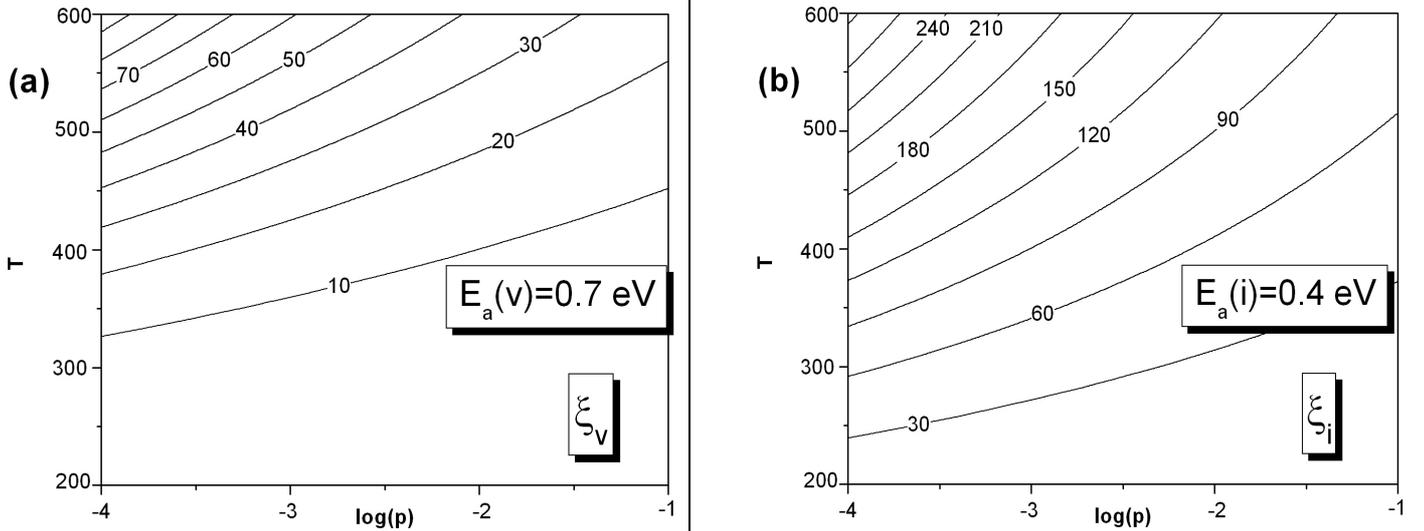

Figure 6. The predicted superlattice constants (in units $a_0$) for vacancy clusters (metal colloids) (a) and interstitial aggregates (halogen gas bubbles) (b) in electron irradiated $CaF_2$ as a function of the dose rate $p$ and the temperature $T$.

The expected superlattice parameters for vacancies and interstitials are plotted in Fig. 6. If we accept $\log p = -3...-2$ and $T = 300$ [8], the expected interstitial superlattice parameter is 30-60 $a_0$, in the perfect agreement with the experimental value of 50 $a_0$ [8]. On the other hand, the expected lattice parameter for vacancies (metallic colloids) is very small, less than 10 $a_0$. In other words, small (and probably, hardly observable) metallic colloids are supposed to adjust to the superlattice of larger halogen bubbles. Our pattern is supported by the conclusion [23] that the observed voids contain fluorine gas.

## 5. Conclusions

We suggested simple relations for the void lattice parameter dependence on the dose rate and the temperature which are in good agreement with basic experimental observations in both metals and insulators. We have shown also that when the ordering of randomly distributed voids occurs, the system "density" (void number per volume) does not change considerably which reminds solid crystallization from melt. In the particular case of $CaF_2$, we reproduced the experimentally observed size of voids, confirmed that these are gas bubbles and predicted existence of small Ca colloids hidden in the void superstructure. The results of relevant kMC modeling of void formation kinetics will be published elsewhere [17]. To our opinion, the secong stage of the process-- void lattice formation -- is a self-organized process in an open dissipative system far from equilibrium which is controlled by defect density fluctuations. *This means that there is no guarantee that the same void superstructure will be reproduced in different (real or computer) experiments under the same conditions.* Thus, the main question which could be raised is: under which conditions there is a chance to observe void superlattices.

**Acknowledgements**
Authors are indebted to A.M. Stoneham and W. Frank for discussions. This work is partially supported by the US DOE BES under grant # DE-FG02 -02ER46005 and Latvia-EURATOM Fission project.

## References

* Electronic address: kotomin@latnet. lv

[1] W. Jaeger and H. Trinkaus, J. Nucl. Mater. **205,** 394 (1993).

[2] A. M. Stoneham, Rep. Prog. Phys., **70**, 1055 (2007).

[3] H.C. Yu and W. Lu, Acta Mater. **53**, 1799 (2005).

[4] J.R. Willis and R. Bullogh, J.Nucl. Mater. **32**, 76 (1969).

[5] J.R. Willis, J. Mech. Phys. Solids, **23**, 129 (1975).

[6] A.M. Stoneham, J. Phys. F. **1**, 778 (1975).

[7] N.M. Ghoniem, D. Waldgraef, and S.J. Zinkle, J. Comp. Aided. Mater. Design, **8**, 1 (2002).

[8] T.H. Ding, S. Zhu, and L.M. Wang, Microsc Microanal **11** (Suppl 2), 2064 (2005).

[9] E. A. Kotomin and V. N. Kuzovkov, *Modern Aspects of Diffusion-Controlled Reactions: Cooperative Phenomena in Bi-molecular Processes,* Vol. **34** of *Comprehensive Chemical Kinetics* (Elsevier, North Holland, Amsterdam, 1996).

V. Kuzovkov and E. Kotomin, Rep. Prog. Phys. **51,** 1479 (1988).

E. Kotomin and V. Kuzovkov, Rep. Prog. Phys. **55,** 2079 (1992).

[10] R. Salazar, A. P. J. Jansen, and V. N. Kuzovkov, Phys. Rev. E **69**, 031604 (2004).

[11] H.L. Heinisch and B.N. Singh, Phil. Mag. **83**, 2661 (2003).

[12] J.H. Evans, Phil. Mag. **85**, 1177 (2005).

[13] K. Krishan, Radiat. Eff. Def. Solids, **66**, 121 (1982).

[14] E.A. Kotomin, V.N. Kuzovkov, M. Zaiser and W.Soppe, Rad.Eff. Def. Solids, **136**, 209 (1995).

[15] E.A. Kotomin, V.N. Kuzovkov, Physica Scripta **50**, 720 (1994).

[16] A.A. Semenov and C.H. Woo, Phys. Rev. B, **71**, 054109 (2005);

A.A. Semenov, C.H. Woo and W. Frank, Applied Phys. Lett. A (2009), in press.

[17] V. Kuzovkov, G. Zvejnieks, E.A. Kotomin, K.D. Li, L.M. Wang, (2009) to be submitted

[18] E.A. Kotomin, V.N. Kuzovkov, G. Zvejnieks, Yu. Zhukovskii, D. Fuks, S. Dorfman, and A.M. Stoneham, Sol. State Comm., **125,** 463 (2003).

[19] C. T. Campbell, Surf.Sci. Reports, **27**, 67 (1997).

[20] T.R.Allen and G.S. Was, Chapter 6 in *Radiation Effects in Solids*, NATO series in science, vol. **235** (Kluwer: eds. K.Sikafus, E.A. Kotomin, B. Uberuaga, 2007)

[21] W. Hayes and A.M. Stoneham, *Defects and Defect Processes in Non-Metallic Solids*, (Wiley, New York, 1985).

[22] V.I. Dubinko, A.A. Turkin, A.S. Sugonyako, D.I. Vainstein, and H.W. den Hartog , phys. stat. sol. (c) 2, **438** (2005) and references therein.

[23] T.H. Ding, Q.M. Wei, K.D. Li, S. Zhu, L. Hobbs, L.M. Wang , to be submitted.

[24] K.D. Li, H.Y. Xiao, and L.M. Wang, Nucl. Inst. Meth. B **266**, 2698 (2008).

[25] S.C. Keeton and W.D. Wilson, Phys. Rev. **B7**, 834 (1973).

[26] K. Atobe, J. Chem. Phys. **71**, 2588 (1979).

[27] M. Huisinga, N. Bouchaala, R.Bennewitz, E.A. Kotomin, V.N. Kuzovkov, and M.Reichling, Nucl. Inst. Meth. **B 141**, 79 (1998).